# Multifractal Solar EUV Intensity Fluctuations and their Implications for Coronal Heating Models

A.C. Cadavid[1], Y.J. Rivera[2], J.K. Lawrence[1], D. J. Christian[1], P.J. Jennings[3], and A.F. Rappazzo[4]

[1] Department of Physics and Astronomy, California State University Northridge,
18111 Nordhoff Street, Northridge, CA 91330, USA;  ana.cadavid@csun.edu
[2] Department of Climate and Space Sciences, University of Michigan,
Ann Arbor, Michigan 48109-2143, USA
[3] 5174 S. Slauson Avenue, Culver City, CA 90230, USA
[4] Department of Earth, Planetary and Space Sciences,
University of California Los Angeles, Los Angeles, CA 90095, USA

Version October 20, 2016

**Abstract**

We investigate the  scaling properties of the long-range temporal evolution and intermittency of Atmospheric Image Assembly/*Solar Dynamics Observatory*  intensity observations in four solar environments: an active region core, a weak emission region, and  two core loops.  We use two approaches:  the probability distribution function  (PDF) of  time series increments  and multifractal detrended  fluctuation analysis (MF-DFA). Noise taints the results, so we focus on the 171 Å waveband , which has the highest signal-to-noise ratio. The lags between pairs of wavebands distinguish  between coronal versus  transition region (TR)  emission. In all physical regions studied, scaling in the range  of 15-45 minutes  is multifractal, and the time series are anti-persistent on the average.  The degree of anti-correlation in the TR time series is greater than for coronal emission. The multifractality stems from long-term correlations in the data rather than the wide distribution of intensities.  Observations  in  the 335 Å waveband can be  described in terms of a multifractal with added noise.   The multiscaling of the extreme-ultraviolet EUV data agrees qualitatively with the radiance  from a  phenomenological model  of impulsive bursts plus noise,  and also  from ohmic dissipation in a reduced magnetohydrodynamics  (RMHD) model  for  coronal loop heating. The parameter space must be further explored to seek quantitative agreement.  Thus, the  observational "signatures"  obtained by the  combined tests of the  PDF of increments and the MF-DFA  offer strong constraints that can systematically discriminate among models for coronal heating.

1.Introduction

Much progress has been made in recent years in understanding aspects of coronal heating. Although a unifying explanation remains elusive, increasingly, observations and numerical simulations indicate the importance of impulsive energy releases. A framework is provided by the nanoflare  model in which generic, small impulsive heating events  occur at  sub-resolution scales  (Klimchuk 2015).  Because it is not possible to use direct observations to probe these minute physical processes, indirect analyses are needed to uncover consequences or "signatures" of the underlying phenomena.

One major approach has been to investigate the scaling of the emission measure as a function of temperature. This means to examine the exponent α in EM(T) ~ $T^{\alpha}$, as well as the range of temperatures over which the scaling occurs, in order to constrain the frequency of heating events in the models (Warren et al. 2011; Warren et al.2012; Bradshaw et al. 2012; Cargill 2014). Although the calculated emission measures can have large uncertainties (Guennou et al., 2012), relationships have been established between the flaring frequency, the nature of the plasma heating in coronal loops, and the scaling properties of the emission measure. In particular, Cargill (2014) has found that the range of exponents encountered in the observations for active region (AR) loops can be reproduced with a nanoflare heating model with triangular pulses in which the amplitude of the events is randomly sampled from a powerrlaw distribution and in which there is a relationship between the amplitude and the time between events. These time intervals have values that vary from a few hundred seconds to about 2000 s, and they are shorter than or comparable to the cooling times required in order to prevent the coronal field from relaxing to a potential state.

In a different approach Terzo et al. (2011) investigated the statistical properties of Hinode X-ray intensity signals. After excluding pixels with very weak signals or with transient brightenings, they defined the intensity fluctuations by subtracting a linear fit to the signals and normalizing each pixel by the photon noise estimated as the standard deviation of the signal from the linear fit. They found that the median fluctuations for a collection of time signals in an AR lay at a small negative value. To generate a synthetic time series they added Poisson noise to the average of the light signal at each time step. Subsequently the time signals were perturbed with a sequence of pulses with a characteristic periodicity and exponential decay time. They found that for simulations with comparable intervals and decay times on the order of 360 s the synthetic signals can reproduce the observations with a distribution of medians with a negative offset. They estimated that only ~60% of the negative shift is due to the Poisson noise. The remaining effect is due to the simulated pulses, which portray the effect of a cooling plasma. A similar analysis has been successfully applied by Jess et al. (2014) to identify the signature of nanoflares in $H_{\alpha}$ chromospheric observations made with the Dunn Solar Telescope using the Hydrogen-Alpha Rapid Dynamics camera.

Alternatively Viall & Klimchuk (2012, 2015) have computed the cross-correlations between contemporaneous pairs of coronal *Solar Dynamics Observatory(SDO)*/Atmospheric Image (AIA) extreme-ultraviolet (EUV) light curves in different wavelength bands. These researchers find that in the case of coronal emission the AR pixels, including those in the loop structures as well as the diffuse emission, the configuration of signal time-lags is consistent with the cooling pattern characteristic of an impulsive nanoflare heating process. That is, the highest temperature emission appears and peaks first, followed by the second hottest, and so on to the least hot. The shortest time series investigated in the 2012 paper has a duration of 2 hours. While results are consistent in that work, these types of analysis could be subject to possible errors caused by large-scale evolutionary events, such as those due to restructuring of the magnetic fields.

Using *SDO*/AIA observations of coronal loops in non-flaring AR cores between the sunspots, Cadavid et al. (2014; henceforth C14) found that the time evolution of fluctuations of the loop apex radiative intensity, the temperature, and the electron density, indicate an underlying impulsive heating process compatible with a high-intensity nanoflare storm (Viall & Klimchuk 2011). The fluctuations are characterized by a sequential cooling pattern with the hot Fe XXI and then the Fe XVIII components of the 131 and 94 Å channels, respectively, leading the emission. The Fourier power spectra for the loop intensity time series displayed scaling with exponents in the range $1 < \beta < 3$ indicating that the signals are statistically nonstationary (Mandelbrot & Van Ness 1968).

These observed nonstationarities may arise from external effects, such as long-term trends or externally driven periodic signals, in which case it is important to remove them. On the other hand, if the nonstationarities are an intrinsic component of the fluctuation mechanism, such as long-term correlations, then they have to be examined in detail. The method of detrended fluctuation analysis (DFA) was designed to determine the true scaling properties of a signal by identifying long-term correlations in noisy and nonstationary time series after accounting for external trends (i.e. Peng et al., 1994, Kantelhardt et al. 2001, Hu et al. 2001, Chen et al. 2001). While DFA allows for the calculation of the second moment of the fluctuations it has been found that in order to fully characterize the intermittency in many time signals it is necessary to analyze the properties of the other moments as well. To address this need properly in the present work, we have turned to the method of *multifractal* detrended fluctuation analysis (MF-DFA) which permits a full description of the scaling properties of the nonstationary time series (Castro e Silva & Moreira 1997; Kantelhardt et al. 2002). It has been successfully applied to evaluate the degree of multifractality in time series in diverse fields (eg. Telesca et al. 2004; Kantelhardt et al. 2006; Movahed et al. 2006; Lin & Fu 2008; Zunino et al. 2008). In this work MF-DFA is complemented by investigation of the scaling properties of the distributions of increments of the data time series. These are useful to discriminate between a monofractal signal, whether correlated or uncorrelated, and a multifractal signal (Budaev 2005).

Our goal here is to show how these analysis techniques can be applied to extract statistical information from observational data and how this in turn can be used to investigate models for impulsive coronal heating. To evaluate the multifractal scaling of AR intensity fluctuations in a non-flaring region we used data from the *SDO*/AIA instrument (Lemen, et al. 2012 ) in the Extreme Ultraviolet (EUV) channels. The present study is focused on the cooler 171 Å channel which has the highest signal-to-noise ratio. We also present results for the hotter 335 Å emission to illustrate the effects of a higher level of noise in the analysis. As examples we investigate the multiscaling properties of the EUV signals in an AR core, roughly between the leading and trailing sunspots, a neighboring weak emission region, and two core loops.
A phenomenological model for intensity fluctuations resulting from a superposition of impulsive signals (Pauluhn & Solanki 2007) is found to present scaling characteristics that qualitatively agree with the observations. We also investigate the properties of the average ohmic dissipation in a reduced magnetohydrodynamics (RMHD) model for coronal loop heating (Rappazzo, et al. 2008) and determine which parameters lead to a better match with the observed scaling properties. To gain further insight into the details of the underlying physical mechanisms it is necessary to look for quantitative agreement between the analysis results of observed and simulated intensities. The analysis methods presented here offer the opportunity to define new signatures of impulsive heating in the form of the degree of multifractality of the intensity fluctuations as well as the numerical values of the scaling exponents of the intensity fluctuations. These more comprehensive criteria can then be used to systematically optimize model parameters as has been done in other studies.

The paper is organized as follows. In section 2 we describe the data, the filtering techniques used to identify the stochastic components of the signals, and the application of the aforementioned lag analysis to select pixels with the characteristic coronal cooling pattern. In section 3 we calculate the distributions of the signal increments, which indicate the need to investigate the multiscaling properties. In section 4 we introduce the mathematical details of the MF-DFA method and the practical considerations that arise when this is applied to observational data. We report in section 5 the results of the multifractal analysis for the solar data. Section 6 describes the models studied and the results of the comparable analysis on the simulated data. In section 7 we summarize our findings and further discuss the implication of the results.

## 2. Data and Preliminary Analysis

We use observations from the AIA (Lemen et al. 2012) on board the *SDO* (Pesnell et al. 2012) in the EUV channels. The data have a $0''.6$ (~ 0.44 Mm) pixel scale (spatial resolution of 1.2 arcsec or ~ 0.9 Mm) and a temporal cadence of 12 s. The analysis focuses on NOAA AR 11250 during its first disk passage around the time in which it crosses the solar meridian at S27. We have analyzed a time series with 1350 temporal pixels (4.5 hr) obtained during UT 12:02-16:32 on 2011 July 13. The images were derotated and aligned. The Heliophysics Events Knowledgebase (http://lmsal.com/hek/hek_isolsearch.html) reports no observations of flares in this AR during the time intervals we are studying. Likewise, no GOES flares are reported during this time period. To investigate the scaling properties we selected two different types of physical locations: a core region between the leading and following sunspots and a neighboring region of weak emission. Figure 1 presents images in the 171, 211 and 335 Å wavebands resulting from an average over the whole time series. As will be shown, the 171 Å signal is most appropriate for the fluctuation analysis since it has by far the highest signal-to-noise ratio. We also study the emission in the 335 Å waveband to explore the effects due to the relative noise contributions. The 211 Å signal is used in the identification of pixels displaying a cooling pattern according to the lag analysis introduced by Viall & Klimchuk (2012).

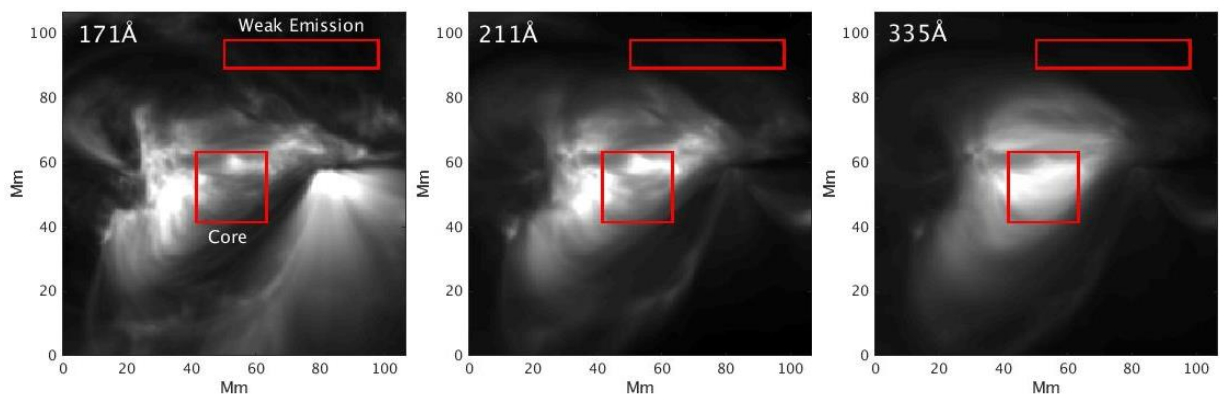

Figure 1. Average intensity of NOAA AR 11250 over the 4.5 hr long time series. The rectangles identify the AR core region (containing 2500 pixels) and the weak emission region (2200 pixels). Left: 171 Å, center: 211 Å, right: 335 Å.

We also report on the properties of two of the AR core loops studied in C14. These were most easily identified by their high variability in the 94 Å emission waveband as compared to a background of steady emission. Since we were not attempting to discover all fluctuating loops in the images we simply adopted a small number of cases by visual inspection of 94 Å animations. Then the loop "spine" was determined by finding the 94 Å emission maxima in each image column, averaged over a few times near the peak. It was found that the 94 Å brightenings tended to be followed in time by brightenings in cooler EUV channels in the progressively cooler sequence 335 Å, 211 Å, 193 Å and 171 Å. Further details can be found in C14. Figure 2 shows two-dimensional projections along the line of sight of the two examples we discuss in this paper. The visible loop brightenings are presented in both the 94 Å and 171 Å wavebands. Loop 1 is defined by 513 pixels and a length 24.8 Mm (Figure 2- top), while loop 2 consists of 693 pixels at a length of 35.5 Mm (Figure 2 - bottom).

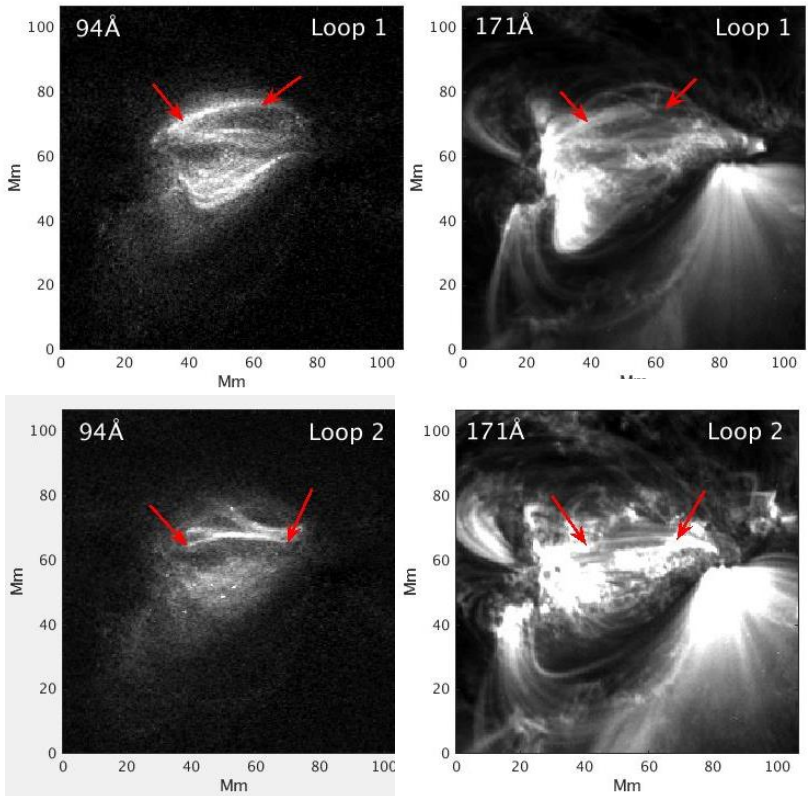

Figure 2. Two-dimensional projections along the line of sight of loops 1 and 2. The arrows indicate the end points. For loop 1 the image in the 94 Å emission (upper left) occurs 40.6 minutes after the start of the time series and the one in the 171 Å waveband (upper right) at 60 minutes. For loop 2, the images selected correspond to the following times: 54.2 minutes for the 94 Å emission (lower left) and 150 minutes for the 171 Å emission (lower right).

To remove macroscopic variations we adopt the technique applied by Terzo et al. (2011) and Jess et al. (2014). The procedure consists of performing a linear fit to the time series for each spatial pixel and then removing those pixels with intensities equal to or greater than some multiple of the best linear fit value at a given time. After trying different threshold levels we find that a factor of two times the best-fit value gives an adequate criterion for excluding transient brightenings. Also, following the aforementioned authors, we eliminate pixels that may be displaying long-term variations caused by the displacement or drift of emitting structures. In this case, we find that for each of the different wavelength bands the percentage of pixels excluded is less than 1% in the core and weak emission regions and 2-4 % in the loops. The total percentage of pixels excluded in each of the signals in the core region is: 171 Å (20%) and 335 Å (2.4%). For the weak region: 171 Å (7.7 %) and 335 Å (13.3%). For the two loops we focus the analysis on the 171 Å signal which, as will be shown, exhibits the sought after scaling properties. In loops 1 and 2, a total of 36% and 14% of the pixels are excluded, respectively.

Based on simulations using the Enthalpy Based Thermal Evolution of Loops (EBTEL) hydrodynamic code (Klimchuk, et al. 2008), Viall & Klimchuk (2015) have compared the ways that the corona and transition region (TR) respond to an impulsive heating event. Unlike the corona, which follows the sequential cooling pattern described, they find that in the transition region all wavelength channels respond at the same time. Evaluation of the time lags between the different channels shows that for ARs on the disk there is a substantial percentage of pixels with zero time lag suggesting the presence of TR emission. In contrast, for radiation above the limb, which does not have a TR contribution, the percentage of pixels with zero time lag decreases. In order to address this point we have calculated the time lags for the 335-171 and 211-171 pairs in the four regions being studied. These are selected from the peak value of the cross-correlation between the two signals. The preparation of the data to eliminate macroscopic excursions ameliorates the possible errors in the lag method caused by large-scale evolutionary events. Zero lags are defined at the time-pixel resolution (12 s) and non-zero lags above or below this threshold. As in Viall & Klimchuk (2012) although a particular time signal goes through more than one heating and cooling cycle, the overall tendency is to present a cooling pattern with emission in the hotter waveband leading. For example in the 335-171 pair in the core region we find the following percentages of pixels for the different lags: 53 % (positive), 14% (negative) and 30% (zero). The corresponding values for the core region 211-171 pair are: 51% (positive), 6% (negative) and 43 % (zero). Similarly in the weak emission region we find the following values: for the 335-171 pair, 76 % (positive), 23 % (negative) and 1 % (zero), and the 211-171 pair, 61 % (positive), 37 % (negative) and 2 % (zero). To better select the pixels with a sequential cooling pattern characteristic of the corona we classify as having "positive lag" the class of pixels which after passing the filtering tests have positive lags in both the 335-171 and 211-171 pairs. The "zero lag" class is similarly defined but in this case both pairs have lags at the time pixel resolution. Table 1 presents the percentage of pixels that survive these tests in each of the regions selected for further study as well as the average time lags. We are reassured in the soundness of the pixel selection by the fact that the average lag for the 335-171 pair is larger than for the 211-171 pair as expected in the sequential cooling pattern. Indeed as shown later in the paper we will find different scaling properties for the coronal and the TR emissions.

## 3. The Probability Distribution Function (PDF) of the Increments

As described in the introduction, the analysis pursued in the present paper was initially motivated by the discovery that power spectra for fluctuating EUV intensity signals in coronal loops presented scaling exponents with values greater than 1, indicating an underlying nonstationary process. Since then Ireland et al. (2015) have also reported spectral exponents larger than 1 for different physical regions using AIA EUV observations in the 171 and 193 Å wavebands. These values of the spectral scaling exponents open the possibility that the observed time series can have the properties of fractional Brownian motion (fBm) (Mandelbrot & Van Ness 1968). While the increments of traditional Brownian motion are uncorrelated, those for the more general fBm are correlated. In these cases the time series of increments, which are identified as fractional Gaussian noise (fGn), satisfy Gaussian probability distributions (Molz et al. 1997).

Given a time signal $x(k)$ the increments at scale $l$ are defined as $y(k) = \delta_l x = (x(k+l) - x(l))$. When the probability distribution functions (PDFs) at scale $l$ and scale $\lambda l$ satisfy the following relation

$$P_l(\delta x) = \lambda^H P_{\lambda l}(\lambda^H \delta x) \tag{1}$$

it is said that the process is monofractal characterized by the Hurst exponent $H$. A value of $H = 0.5$ indicates an uncorrelated series, $H > 0.5$ corresponds to a long-term-correlated ("persistent") series, and $H < 0.5$ corresponds to an "anti-persistent" series. For more complex "multifractal" signals, as will be shown, many exponents are required to characterize the scaling properties (Budaev 2005; Kiyani et al. 2007). A convenient way to discriminate between these two types of processes is by calculating the scaled probability distribution function (PDF) defined as:

$$P_s(y) = \frac{1}{\sigma} P\left(\frac{y - <y>}{\sigma}\right) \tag{2}$$

where $\sigma$ is the standard deviation and $P$ is a histogram of the scaled increments. For monofractal processes, such as white noise or Brownian motion, that are uncorrelated, or for correlated fBm, at all scales, the PDFs have a Gaussian profile. For multifractal signals the PDFs are heavy-tailed at small scales and approach a Gaussian at large scales (Budaev 2005).

In order to better characterize the properties of the distributions we will also report the values of the skewness and the excess kurtosis. The skewness gives a measure of the asymmetry of the data about the mean. Positive skewness implies that the data are more spread (it has a longer tail) to positive values and correspondingly negative skewness indicates a longer tail to negative values. A normal distribution has zero skewness. The excess kurtosis provides a measure of how strong are the contributions of the outliers, i. e. the "heaviness" of the tails. If the distribution has a higher (lower) influence of outliers as compared to the normal distribution, the excess kurtosis is greater (less) than zero. In this paper we have calculated these quantities using the routines provided by MATLAB which define them by:

$$\text{skewness} = \left(\frac{1}{n}\sum_{i=1}^{n}(x_i - \bar{x})^3\right)/\sigma^3 \qquad \text{excess kurtosis} + 3 = \left(\frac{1}{n}\sum_{i=1}^{n}(x_i - \bar{x})^4\right)/\sigma^4$$

where $n$ is the number of values, $\bar{x}$ is the mean, and $\sigma$ the standard deviation.

These values give a general characterization of the particular data sample under study drawn from a larger population. However, it is possible that the sample drawn appears skewed or peaked while the underlying population is normal. In order to determine if the values of skewness and kurtosis are statistically significant we have turned to the test statistic by using a Z-value for the particular quantity (Zar 1996; Sheskin 2007). In its simplest form the Z-value is a standardization. For example, given a measurement $x_i$ from a population with mean $\mu$ and standard deviation $\sigma$, the Z-value is given by $Z = (x_i - \mu)/\sigma$. To decide whether to accept or reject the null hypothesis that the population is normal, the Z-value for the measurement is compared to the critical value in a normal distribution. For $Z_{crit} = 1.96$, 95% of the values in a normal distribution lie within $\mu \pm Z_{crit}\sigma$. Then at the 95% confidence level the null hypothesis cannot be rejected if $Z_{measured} < 1.96$. To calculate the Z-values for skewness and kurtosis we have used the formulas originally developed by D'Agostino (1970) and D'Agostino & Stephens (1986), and summarized in pages 116-119 in Zar (1996).

We have found that in most of the distributions presented in figures 3 and 4 the skewness and excess kurtosis values encountered are significant according to the test statistic described. In the following description the few exceptions are highlighted. We must note that according to Tabachnik & Fidell (1996) for large $n$ samples as encountered here ($\sim 10^4 - 10^6$) there can be an incorrect tendency to rule out normality in the test statistic for skewness.

Figure 3 (left) displays the logarithm of the scaled PDF for the increments of core time series in the 171 Å channel corresponding to pixels with positive lag, as defined in Section two. The increment scale increases from top to bottom in the order 2, 8, 32, and 128 time steps corresponding to 0.4, 1.6, 6.4, and 25.6 minutes respectively. We have investigated the distributions for larger increment scales but because of the lower number of realizations the histograms do not represent accurate distributions. The PDF curves are "leptokurtic" (peaked with heavy tails) for small time increments and become "quasi-Gaussian" for large temporal scales. This progression indicates that the underlying signals have multifractal properties. For all scales we find postive skewness with values ranging between 0.02 (1.6 minute) to 0.32 (25.6 minutes). The value of the excess kurtosis decreases for increasing scales as follows: (18, 5.6 , 2.6, 2.6). For the zero lag pixels in the core 171 Å waveband the shapes of the distributions of increments are similar to those of the positive lag case for the first three time scales (Figure 3 – center). For the largest scale it appears more rounded at small values of the rescaled increments. The skewness varies between 0.16 (6.4 minutes) and 0.52 (25.6 minutes). The values of the excess kurtosis are (4.0, 2.5, 1.4 , 1.6). In the case of the 335 Å band for positive lags (Figure 3 – right) the distributions appear to have a Gaussian shape for small values of the scaled increments but with heavy tails as indicated by the large excess kurtosis values (34, 28, 17, 5.0 ). The values of the skewness, except for the 1.6 min increment scale are all negative and statistically significant ranging between $4.8 \times 10^{-2}$ (0.4 minute) and 0.16 (25.6 minutes). For the 1.6 minute case the skewness is positive $3.8 \times 10^{-2}$ and with a Z-value = 1.80 therefore it is not possible to rule out the null hypothesis. Since the signal-to-noise ratio is smaller in the 335 Å waveband these results could be representing the combined effect of the signal influenced by noise with Gaussian statistics.

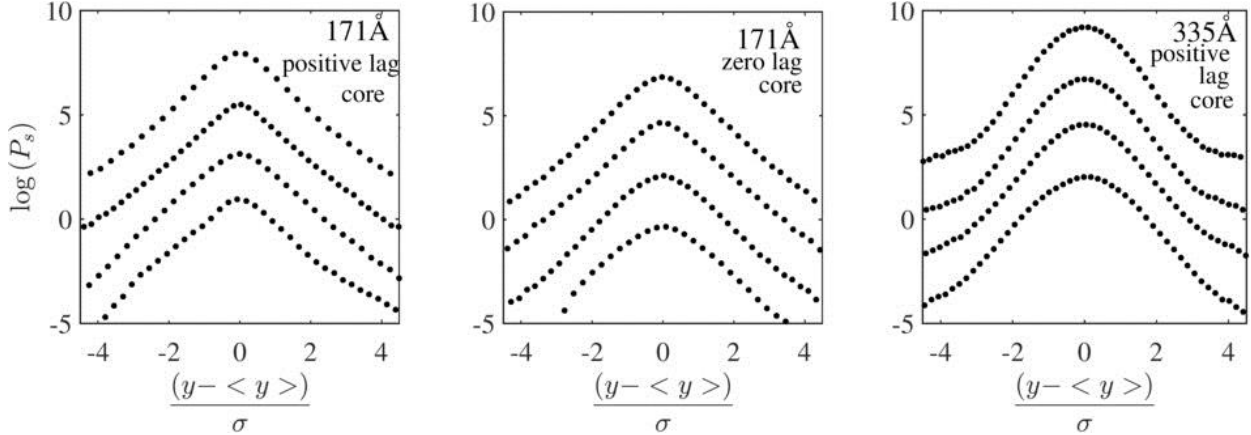

Figure 3. Scaled distributions of increments at increment scales 0.4, 1.6, 6.4 and 25.6 minutes in increasing order from top to bottom. To ease visualization the curves have been shifted downward in units of $\log(P_S)$ relative to the smallest scale curve. Left: core 171 Å for positive lag pixels with vertical shifts (1,5, 3.5, 5.0). Center: core 171 Å for zero lag pixels with vertical shifts (1.5, 3.5, 5.5). Right: core 335 Å positive lag pixels and vertical shifts (2.5, 4.5, 6.5).

Figure 4 presents the corresponding results for positive lag pixels in four more cases. For the weak emission (Figure 4 - top row), the PDF curves have a more rounded shape for small values of the scaled increments than is seen for the core region. This is an effect of the noise component in the signals making a more noticeable contribution at small increment sizes (to be further explored in section 5). While the excess kurtosis for small scales still starts at large values, they are comparatively smaller than in the core. For the 171 Å signal we find the following excess kurtosis (5.3, 4.3, 3.1, 1.5), and skewness (-0.02, -0.03, 0.08, 0.14). All values are determined to be significant. In the case of the 335 Å waveband (Figure 4 – top right) the self-similarity of the curves at different scales is apparent. There is still some evidence of a small deviation from the Gaussian form both in the shape of the curves and in the values of the excess kurtosis: (1.1, 1.0, 1.0, 0.9). The skewness values are $(0.26, -0.44, -2.43, -7.02) \times 10^{-2}$. All values are statistically significant except for the skewness at a time scale of 0.4 min which has a Z-value=1.21. Again the small skewness and variations in its sign may be a result of the noise contribution. The degree of multifractality is diminished in comparison to the 171 Å example. As we will show in section five this is a consequence of the noise component dominating these data. The bottom row in Figure 4 displays the results for the two loops for positive lag pixels and emission in the 171 Å waveband. We encounter a progression of curves with similar characteristics to those of the core region 171 Å emission. For loop 1 using the same temporal scales the values for the excess kurtosis are (12, 4.4, 2.1, 1,3 ), and the skewness varies from 0.09 (1.6 minute) to 0.35 (25.6) minutes. In the case of loop 2 the skewness varies between 0.15 (25.6 minutes) to 0.61 (0.4 minute) and the excess kurtosis is: (20, 8.9, 3.6, 0.4). Finally, for loop 2 but for zero lag pixels, we found excess kurtosis (6.9, 2.3, 0.7, 0.1) and negative skewness ranging between −0.07 (0.4 minute) and −0.191 (6.4 minutes).

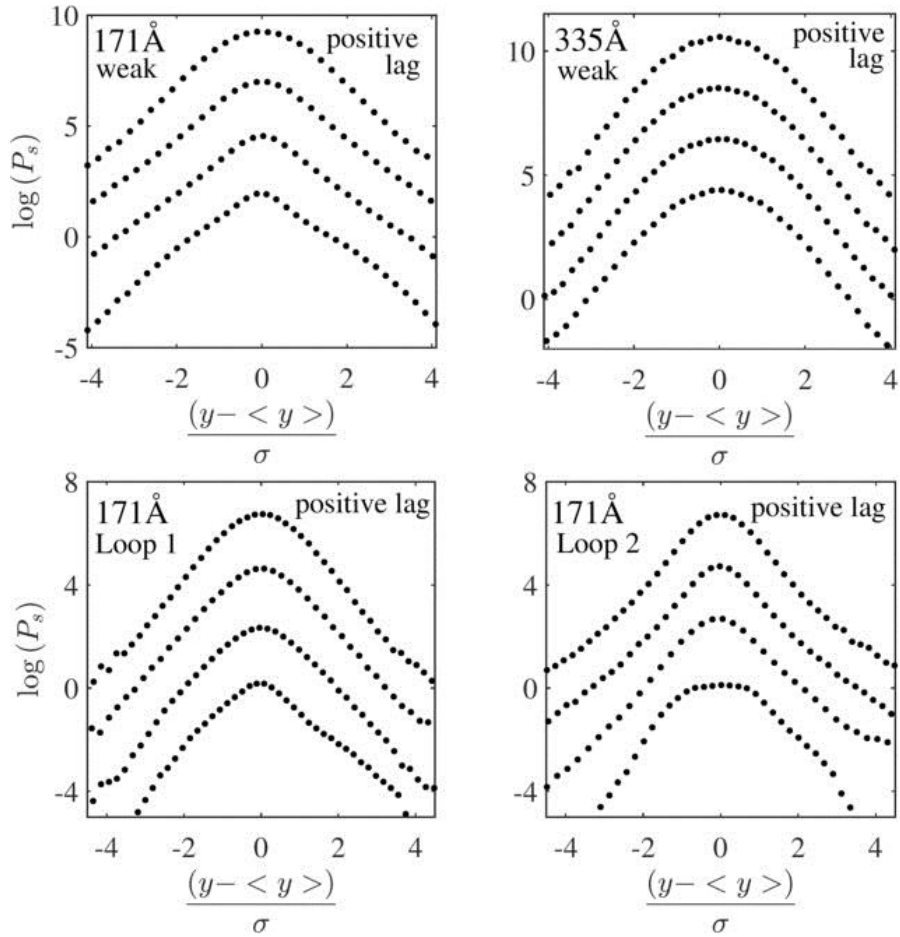

Figure 4. Scaled distribution of increments for positive lag pixels. The increment scales are 0.4, 1.6, 6.4, and 25.6 minutes in increasing order from top to bottom. To ease visualization the curves have been shifted downward in units of $\log(P_s)$ relative to the smallest scale curve. The shifts are: (2, 4, 6) for the 171 Å weak emission (top-left), (2, 4, 6) for 335 Å weak emission (top-right), (1.5, 3, 4) for loop 1 (bottom-left) and (1.5, 3, 4.5) for loop 2 (bottom-right).

Since we find both positive and negative skewness and the values are small we return to the warning by Tabachnik & Fidell (1996) about the possibility of having erroneously found statistical significance and that normality cannot be ruled out. So additional considerations must be introduced to be able to extract information from the results encountered. Except for the loop 2 zero lag case, the negative skewness is found in regions where the noise may have a stronger contribution. If indeed this is the case, the positive skewness found in all the other regions is interesting since, as will be shown, these characteristics can be related to the properties of the phenomenological model for impulsive heating.

## 4. Multifractal Detrended Fluctuation Analysis

As shown in the previous section the distribution of increments for coronal EUV emission have properties that suggest multifractality. To test whether this is indeed the case, in this paper we are going to follow the method of multifractal detrended fluctuation analysis (MF-DFA) as introduced by Kantelhardt and collaborators (eg. Kantelhardt et al. 2002; Kantelhardt et al. 2006; Kantelhardt 2008). We will discuss ways to account for a spurious multifractality effect in MF-DFA at the end of this section.

Analyses of time series in diverse fields present complex examples in which the scaling properties cannot be described by a single scaling exponent, as could be done for a monofractal. In some cases the scaling exponents and correspondingly the type of long range temporal correlations can be different at small temporal scales as compared to large temporal scales. In more complex cases time series may actually contain "many interwoven fractal subsets" (Kantelhardt et al. 2002, page 88) and a full description requires knowledge of multiple scaling exponents for the higher (and lower) moments of the variables. In these cases the multifractal property is manifested by the presence of different scaling exponents, and therefore different long-term correlation properties, for small fluctuations than for large fluctuations. The multifractality in a time series can have two main origins. It can result from intermittency in the time series which is characterized by a broad probability distribution, with fat tails, of its values, or it can be a consequence of long-term correlations. To test this property the scaling analysis can be applied to a shuffled version of the time series. If the multifractality is a result of long-term correlations it will disappear for the shuffled series. When both types of multifractality are relevant the shuffled series will show a decrease in the degree of multifractality compared to that of the original time series (Kantelhardt et al. 2002). Below, following Kantelhardt al. (2002), we present a schematic description of the steps followed in the MF-DFA.

Given a time series $x(k)$ of length $N$, the accumulated times series or "profile" $Y$ is defined by

$$Y(i) \equiv \sum_{k=1}^{i} [x(k) - <x>], \quad \text{for} \quad i = 1, \ldots, N \qquad (3)$$

By performing the scaling analysis on the cumulative sum there is no need to consider a priori whether the time series is stationary or nonstationary (Hardstone et al. 2012). The profile is then divided into $N_s = \text{int}(N/s)$ non-overlapping segments of length (scale) $s$. Typically, $N$ is not an exact multiple of $s$, so to account for missing pieces at the end of the series, the procedure is repeated by deriving the profile in reverse, starting from the end of the series. In total, then, there are $2N_s$ segments at each scale.

The next step consists of calculating a least squares fit of a polynomial $y_\nu$ to each segment $\nu$. The fitted polynomial is then subtracted from each segment of the profile and the variance is calculated by

$$F^2(v,s) \equiv \frac{1}{s}\sum_{i=1}^{s}\left(Y[(v-1)s+i]-y_v(i)\right)^2 \tag{4}$$

for $v = 1,\ldots,N_s$. The procedure is repeated for the second case in which the series is divided into segments starting at the end:

$$F^2(v,s) \equiv \frac{1}{s}\sum_{i=1}^{s}\left(Y[N-(v-N_s)s+i]-y_v(i)\right)^2 \tag{5}$$

for $v = N_s+1,\ldots,2N_s$.

The fitting polynomial can be of linear, quadratic, or higher order. The $m$th order MF-DFAm subtracts $m$ order trends in the profile and $m$-1 order trends in the original time series.

Finally an average over all segments is performed to obtain the $q$th order fluctuation function

$$F_q(s) = \left(\frac{1}{2N_s}\sum_{v=1}^{2N_s}[F^2(v,s)]^{q/2}\right)^{1/q} \tag{6}$$

where $q \neq 0$. For $q = 0$ the expression becomes:

$$F_0(s) = \exp\left(\frac{1}{4N_s}\sum_{v=1}^{2N_s}\ln[F^2(v,s)]\right) \tag{7}$$

If the time series is long-term power law correlated $F_q(s)$ satisfies a scaling law:

$$F_q(s) \sim s^{h(q)} \tag{8}$$

For large scales $s > N/4$ the results become statistically unreliable (Kantelhardt et al., 2002) so in practice the fluctuation functions for the data are only calculated up to that scale. Also for small time scales the method can present deviations from the scaling law so in many cases the first 10 temporal pixels are ignored. The function $h(q)$ is identified as the "generalized Hurst exponent" obtained from log-log plots of $F_q(s)$ vs $s$ for the appropriate scaling range. For stationary series, including fGn, $0 < h(2) < 1$ and $h(2) = H$ the Hurst exponent. It can be shown for fBm, with $1 < h(2) < 2$, that $H = h(2)-1$ (Heneghan & McDarby 2000). Following Yu et al. (2011), in the present study we apply these relations between $h(2)$ and $H$ for the general case. For a monofractal time series $h(q)$ is constant and independent of $q$ (however, see discussion in the next paragraph). For a "textbook" multifractal, $h(q)$ is a non-increasing function of $q$ that tends to constant values asymptotically for both positive and negative moments. Therefore a standard measure of the degree of multifractality is given by $\Delta h = h(q_{min})-h(q_{max})$. An ideal monofractal would give $\Delta h = 0$. In practice this condition is applied for relative comparisons even if the fully asymptotic values for the generalized exponents have not been reached. For large fluctuations the contributions from positive $q$ moments dominate the sum in Equation (6). Similarly for small fluctuations the largest contributions come from the negative $q$ values. Therefore the scaling properties for segments with large (small) fluctuations are described by $h(q)$ for positive (negative) $q$. Because of this when noise with small amplitude is added to a signal its presence primarily affects the generalized exponents on the negative $q$ side. To make contact with other approaches to multifractal analysis Kantelhardt, et al., (2002) show that the generalized Hurst exponent $h(q)$ is directly related to the Renyi (or mass) exponent $\tau(q)$ (Feder, 1988) by $\tau(q) = qh(q) - 1$.

While MF-DFA can give more accurate results than well-known alternative methods (Oświecimka et al. 2006), still the finite length of the data string used causes a spurious multifractal effect. Grech & Pamula (2013) have performed an extensive study to investigate this problem by applying MF-DFA to generated data of persistent series with Hurst exponents $0.5 < H < 1$ and a corresponding autocorrelation coefficient $\gamma = 2(1 - H)$. Although the time series are expected to be monofractal by construction it is found that $\Delta h \neq 0$. The strength of this spurious multifractality is linear with respect to the autocorrelation exponent of the data and decays as a power law as a function of the length of the time series. In consequence it affects shorter time series and those authors recommend that the MF-DFA must be applied to data with at least $2^{10}$ pixels. Even for very large time series the spurious multifractality, although small, it is still present. For example, for 100 series of length $2^{16}$ pixels of uncorrelated noise the degree of multifractality encountered is 0.06. To get an estimate of this finite size effect (FSE) for the observational data we calculate the spurious multifractality strength for realizations of generated time series of uncorrelated noise with the same length as the observed time series. We define a "corrected" multifractality strength by subtracting this spurious quantity from the $\Delta h$ measured for the data.

## 5. Results for the MF-DFA of the EUV Coronal Intensities

In the present work the C code developed by Goldberger et al. (2000) to perform the detrended fluctuation analysis (for $q = 2$) has been modified to include all moments. The results have been cross-checked with a slower code developed in MATLAB (Ihlen 2012). We have investigated the dependence of the fluctuation functions on the order of the detrending polynomial and find that $m = 2$ in which a linear trend is subtracted from the data, in general leads to the best scaling properties. We explore moments in the range $-20 \leq q \leq 20$ with $\Delta q = 0.5$. For the time series with 1350 pixels the maximum scale calculated is about one-fourth or 323 pixels=65 minutes.

Before calculating the moments of a general order $q$, it is instructive to investigate the scaling properties of the second moment $q = 2$ since this exponent provides information on the temporal correlations. In the case of fBm and fGn signals it has been found that there is a relation between the spectral exponent $\beta$ and the Hurst exponent given by $\beta = 2H + 1$ (Heneghan & Mc Darby, 2000). In the case of a multifractal signal no such relation has been established between $\beta$ and $h(2)$, but as we embark on this analysis we will keep in mind the information provided by the power spectra. We find that for the core region emission in the 171 Å there is a small amount of Fourier excess power in the range 2.5-14 minutes which covers the acoustic regime. This is compatible with the range of 1-10 mHz found by Ireland et al. (2015) in particular for signals at loop footpoints and moss. These authors suggest that this excess emission could be due to contributions from below the corona as described by De Pontieu et al. (2005). There is no evidence of excess emission in the spectra for the weak emission region in the 171 and the 335 Å wavebands. Figure 5 (left) presents the average second order fluctuation function $F_2$ as a function of timescale $s$ for positive lag pixels (associated with coronal emission) in the core 171 Å waveband. To determine the large timescale scaling properties we have selected the best fitting range above the enhanced emission limit: 14.8 – 32.2 minutes. We find a generalized Hurst exponent $h(2) = 1.26 \pm 0.03$ (Figure 5 – left) as reported in Table 1. In the case of the zero lag pixels (associated with TR emission) the generalized exponent for large time scales is $1.14 \pm 0.02$ (Figure 5 – center). In this case the scaling range can be extended to 14.8-41.8 minutes. As described in Section 4 the Hurst exponent which characterizes the degree of correlation of the time series is given by $H = h(2) - 1$, for nonstationary time series with $h(2) > 1$, and $H = h(2)$ if $h(2) < 1$. If indeed the type of lag allows us to distinguish between the coronal vs the TR contributions we are finding that the increments in TR signals have a stronger degree of anti-correlation ($H = h(2) - 1 = 0.14$) than do those in the corona ($H = h(2) - 1 = 0.26$). In both cases the values of the generalized Hurst exponents indicate that the underlying processes are nonstationary and anti-persistent with $H < 0.5$, which implies that for "large" temporal scales positive fluctuations tend to be followed by negative fluctuations and vice versa.

Figure 5 (right) displays the second order fluctuation function for positive lag pixels in the core 335 Å waveband. In this case we proceed to calculate the scaling exponents for both small and large temporal scales. For a scaling range 16.2 – 41.8 minutes the scaling exponent is $h(2) = 1.29 \pm 0.03$ which, as for the 171 Å, case corresponds to an anti-persistent signal with $H = h(2) - 1 = 0.29 < 0.5$. For small temporal scales the fit in the range 1.6 - 3.2 minutes is $H = h(2) = 0.62 \pm 0.01$ tending towards the value for an uncorrelated time signal. The cross-over time defined as the intercept of the two tangents is 13.6 minutes. At large temporal scales the correlation properties are qualitatively the same as in the case of the cooler channel, but, as we will show in the next section, the reduced persistence at small scales can be attributed to some type of noise contribution to the signal. The signal to noise ratio is higher in the 171 Å than the 335 Å waveband. Thus in the latter case it is expected that the noise component is affecting the value of the scaling exponent at small time scales and that its "true" value is larger than what is observed. This implies that the noise-free positive correlation at small scales would be stronger than it appears to be here.

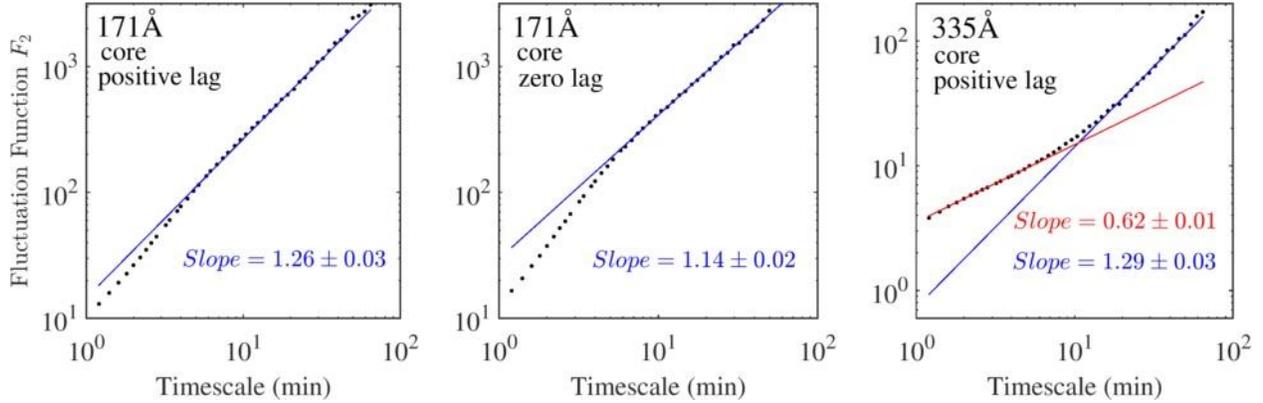

Figure 5. Average second moment fluctuation function $F_2$ for signals in the core region. Left: positive lag pixels and signal in the 171 Å waveband. Center: zero lag pixels and signal in the 171 Å. Right: positive lag pixels and signal in the 335 Å waveband. The linear fits at large temporal scales and corresponding slopes $h(2)$ are presented in each case. For the 335 Å case the tangent line for small temporal scales is also included.

As we have seen the scaling properties at small scales are strongly affected by contributions from the noise component in the signals. Thus, in order to investigate the true multiscaling properties of the data we focus our analysis on the large temporal scales, where the effect of noise is weaker. Figure 6 (top row) presents the average general fluctuation functions $F_q(s)$ for the core region in the 171 Å waveband for pixels with positive lag (Figure 6 – left) and zero lag (Figure 6 – center). The right column presents the results for the positive lag pixels in the 335 Å waveband. Only selected values of the moments $q$ are displayed so as to ease visualization. The variation in the fluctuation functions is similar for the 171 Å signals for pixels with positive or zero lag. In the case of the 335 Å signal there is large distortion at small scales and for negative moments. This is a manifestation of the noise component within the signal with its higher contribution at small scales and negative $q$. The bottom row in Figure 6 displays the corresponding generalized Hurst exponents for large temporal scales using the same fit as in the case for $q = 2$. The values for $h(2)$, already introduced, the degree of multifractality, $\Delta h$, and the scaling ranges are summarized in Table 1.

All wavebands show multifractality, but the relatively stronger noise contribution in the 335 Å channel has the effect that the generalized Hurst exponents for negative $q$ are reduced in value while the ones for positive $q$ increase in value, the combined effects leading to a reduction in the degree of multifractality. Figure 4 of Ludescher et al. (2011) presents a useful illustration on how the generalized Hurst exponent varies as noise of increasing relative amplitude is added to a generated multifractal signal leading to $h(q)$ curves as obtained in our case for the 335 Å waveband (Figure 6 – bottom right). Note that a canonical multifractal curve for $h(q)$ would be non-increasing with $q$. Thus the $h(q)$ for the 335 Å case in Figure 6 (bottom-right) shows some deviation from proper multifractal scaling with $h(q)$ increasing for $q$ near zero, as found in Ludescher et al. (2011). Those authors refer to this situation as a case of "corrupted multifractality." The noise effect precludes us from investigating whether the hotter emission channel has different scaling properties from those of the cooler channel, which could represent signatures of a different underlying physical process. This result indicates that a high signal-to-noise is required to accurately diagnose the multifractility of EUV signals spanning a wide range of moment $q$ values.

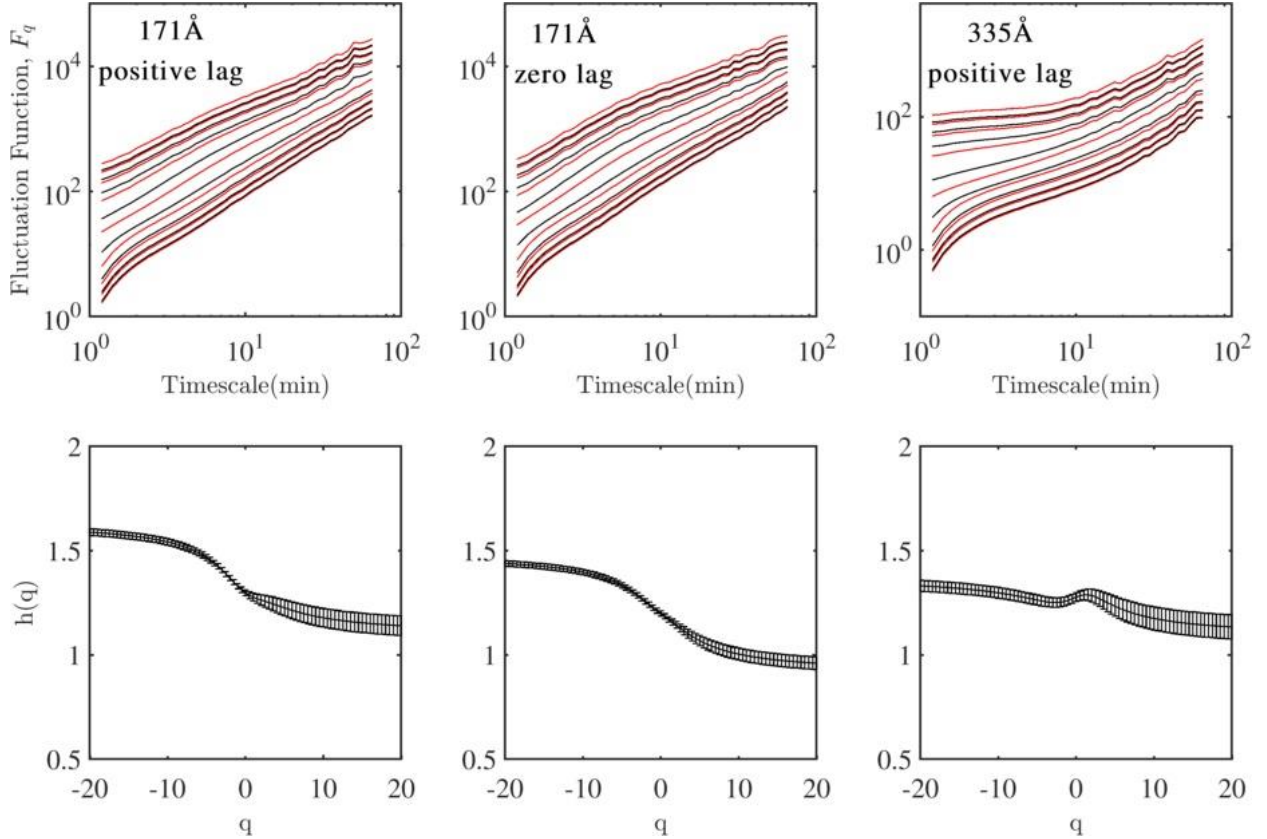

Figure 6. Top row: log-log plot for the average $F_q(s)$ vs scale $s$ for signals in the core region. Left: positive lag pixels and signal in the 171 Å waveband. Center: zero lag pixels and signal in the 171 Å. Right: positive lag pixels and signal in the 335 Å waveband. For clearer visualization not all curves are presented. From top to bottom $q$ varies from 20 to -20 in intervals of 2.5. Bottom row: the corresponding generalized Hurst exponents. Error bars describe the uncertainty in the linear fit to the scaling region (fitting range) in $F_q(s)$.

Based on the number of pixels for the different regions, as an approximation we have applied the MF-DFA procedure to a data set of 2000 cases of generated random Gaussian noise with time series of length equal to the observational data (1350 time pixels). In this case we find spurious multifractality of $\Delta h = 0.16$ due to the finite size effect. We subtract this value from the original Δh values for all time series and calculate the "corrected" multifractality degree presented in Table 1. In the core, for either the pixels associated with TR emission (zero lag) or coronal emission (positive lag), the degree of multifractality in the 171 Å is still significant after taking into account this correction. For the 335 Å case the presence of noise reduces the degree of multifractality to effectively zero.

| Region | Avg. Lag 335-171 (min) | Avg. Lag 211-171 (min) | Percent pixels | Wavelength (Å) | Fit Range (min) | h(2) | Δh | Corrected Δh |
|---|---|---|---|---|---|---|---|---|
| Core | 38.0 | 29.4 | 39 | 171 | 14.8-32.2 | 1.26 ± 0.03 | 0.45 ± 0.05 | 0.29 ± 0.05 |
| | 0 | 0 | 23 | 171 | 14.8-41.8 | 1.14 ± 0.02 | 0.48 ± 0.03 | 0.32 ± 0.03 |
| | 39.9 | 29.1 | 40 | 335 | 16.2-41.8 | 1.29 ± 0.03 | 0.20 ± 0.06 | 0.04 ± 0.06 |
| Weak Emission | 46.3 | 34.9 | 48 | 171 | 14.8-41.8 | 1.37 ± 0.03 | 0.50 ± 0.03 | 0.34 ± 0.03 |
| | 46.9 | 34.8 | 49 | 335 | 14.8-41.8 | 0.78 ± 0.01 | 0.04 ± 0.02 | 0 |
| Loop 1 | 48.9 | 17.2 | 71 | 171 | 14.8-35.2 | 1.41 ± 0.03 | 0.50 ± 0.05 | 0.34 ± 0.05 |
| Loop 2 | 33.7 | 25.8 | 41 | 171 | 14.8-41.8 | 1.34 ± 0.04 | 0.57 ± 0.06 | 0.41 ± 0.06 |
| | 0 | 0 | 17 | 171 | 14.8-41.8 | 1.15± 0.03 | 0.67 ± 0.03 | 0.51 ± 0.03 |

Table 1. Summary of the characteristic average lags for the 335-171 and 211-171 pairs. The "percent of pixels" refers to the cases selected by the condition that both pairs 335-171 and 211-171 have either positive or zero lags. These pixels were used in the analysis which led to the listed scaling ranges, generalized Hurst exponent $h(2)$ and the degree of multifractality Δh.

Analysis of the shuffled time series results in values of $h(2) \sim 0.5$ which correspond to uncorrelated processes. We also find that $\Delta h < 0.18$ for the shuffled time series. Since these values of Δh are on the order of the degree of multifractality for an uncorrelated time series, we conclude that they result from the finite size effect and that, therefore, the observed multifractality is due to the long-term correlations in the time series and not due to heavy tails in the intensity probability distributions.

Figure 7 displays the averaged fluctuation functions and the generalized scaling exponents for the weak emission region for pixels in the positive lag class corresponding to coronal emission. The percentage of pixels with zero lag after the selection process was very small (< 4%). The values of the various parameters are summarized in Table 1. For the 171 Å waveband (Figure 7, bottom left) the Hurst exponent curve is shifted slightly upwards in comparison to the cases for the core region. The higher value of the Hurst exponent ($H = h(2) - 1 = 0.37 \pm 0.03$) still indicates the presence of an anti-persistent process but the degree of anti-correlation is reduced in comparison to the signals in the core region. The corrected degree of multifractality ($0.34 \pm 0.03$) is comparable to that found in the core ($0.29 \pm 0.05$). Analysis of the shuffled time series indicates that the multifractality is caused by the long-term correlations at large temporal scales. In the case of the 335 Å waveband the multifractality has disappeared (Figure 7, bottom right), and as the relative noise component increases the Hurst exponent settles into a value ($H = h(2) = 0.78 \pm 0.01$) which implies that the increments of the time series have positive correlation. As shown in Figure 4 (top right) for small values of the rescaled increments the PDFs tend toward a Gaussian shape for all temporal scales. This result, together with the value of $H$ taken together, suggests that the noise in the signal belongs to the broad class of fractional Gaussian noise (fGn) introduced by Mandelbrot & Van Ness (1968). It is traditionally assumed that the noise contribution in the intensity observations satisfies either Gaussian or Poisson statistics. However the present result suggests that this background emission has other properties. This is compatible with the PDFs of increments which showed self-similarity corresponding to a monofractal but with excess kurtosis values around 1.

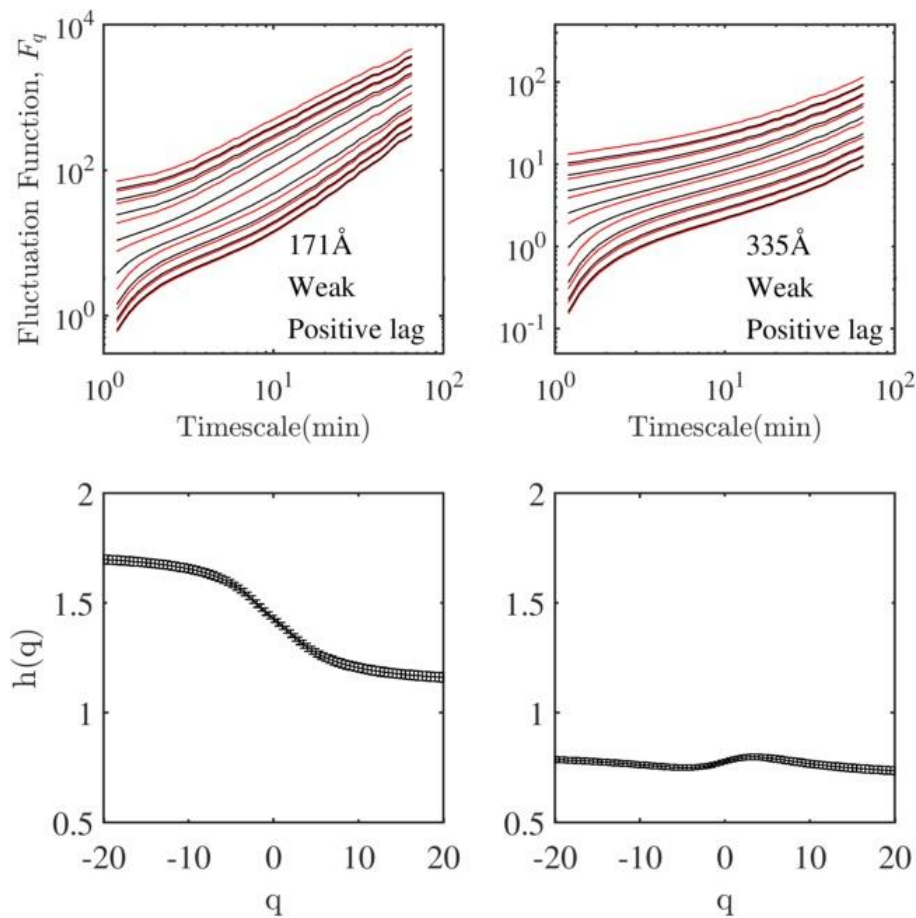

Figure 7. Top row: log-log plot for $F_s(q)$ vs scale in the weak emission region for pixels in the positive lag class. Left: 171 Å waveband. Right: 335 Å waveband. For clearer visualization not all curves are presented. From top to bottom $q$ varies from 20 to -20 in intervals of 2.5. Bottom row: the corresponding generalized Hurst exponents.

Since we have shown that the 335 Å waveband signals are "contaminated" by noise, for the loops we only consider the 171 Å emission. For both loops Figure 8 presents the second moment fluctuation functions (Figure 8, top row) and the generalized Hurst exponents (Figure 8, bottom row) for pixels with positive lag. We also show the corresponding results for zero lag pixels in loop 2 (right column). In the case of loop 1 the percentage of pixels with zero lag, satisfying the selection criteria, was very small (< 1%) and this case is not considered. For scaling ranges 14.8-35.2 minutes (loop 1) and 14.8 – 41.8 minutes (loop 2) the two loops present comparable multifractality degrees with corrected values of 0.34 and 0.41. The Hurst exponents are close ($H = h(2)-1 = 0.41$ for loop 1 and 0.34 for loop 2) implying an anti-correlated process for large temporal scales. For the case of the zero lag pixels in loop 2 the corrected degree of multifractality is 0.51, higher than in any of the other cases studied. The Hurst exponent is ($H = h(2)-1 = 0.15$) comparable to the value of 0.14 encountered for the zero lag pixels in the core region. The shuffled time series for the positive lag pixels in both loops 1 and 2 have degrees of multifractality with values within the one attributed to the FSE ($\Delta h = 0.16$). In the case of the zero lag pixels for loop 2 we find ($\Delta h = 0.18 \pm 0.01$). This leads to a corrected degree of multifractality which is very small but still non-zero, indicating that a small degree of the multifractality is caused by large values in the distribution of intensities. This might explain why the corrected degree of multifractality for this region (0.51) is larger than for the other cases.

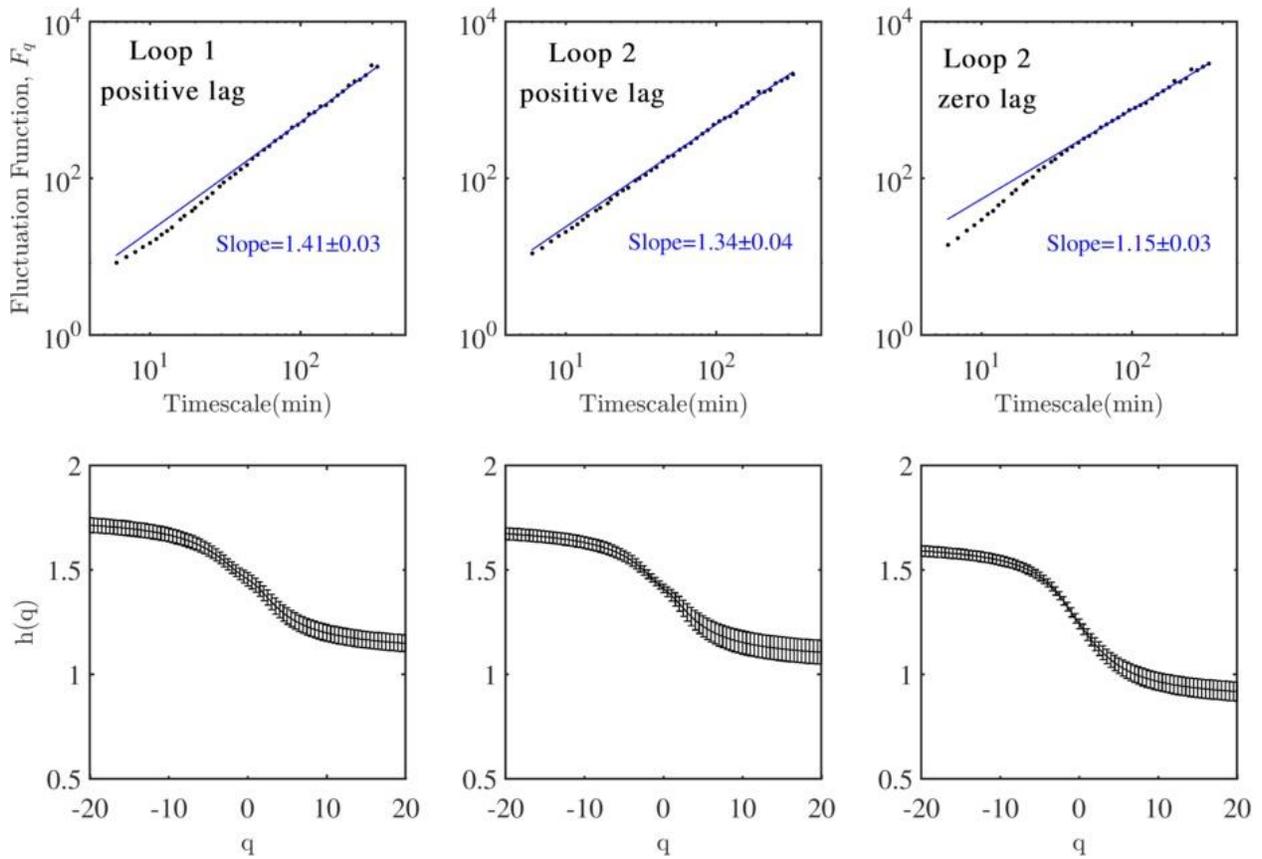

Figure 8. Average second moment fluctuation function $F_2$ (top row) and corresponding generalized Hurst exponents (bottom row) for 171 Å signals in the loops. Left: positive lag pixels for loop 1. Center: positive lag pixels for loop 2. Right: zero lag pixels for loop 2. The linear fits at large temporal scales and corresponding slopes $h(2)$ are presented in each case.

We note that in all the results presented the upper limit for the scaling range appears to be determined by the duration of the data strings and the restrictions of the MF-DFA. There is the potential that by using longer data strings, the scaling range can be extended to larger scales. In particular Auchére et al. (2014) have used time series in the 195 Å waveband of EIT and found that average spectra for both quiet sun and active regions present power laws in the approximate range 20 minutes -20 hr.

## 6. Comparison to Modeled Data

### 6.1 Simple Model for Impulsive Heating

To understand the different forms of scaling encountered in the previous sections, we have made use of a technique devised by Pauluhn & Solanki (2007). This was introduced specifically to relate modeled nanoflaring to the probability distribution function and the power spectra of observed radiances in the quiet Sun TR and lower corona. Here we will apply this idea informally as a kind of toy model that we manipulate to aid in interpreting the different scaling patterns we find for data in different AIA wavelength bands or emitted under different physical circumstances.

The method is based on the idea that an observed intensity time series is caused by a sequence of "nanoflares" that occur on the Sun at each observational time step with some probability $P_f$. The intensity $I_0$ of the pulse at its origin point lies in a fixed range, and the actual value in that range is selected according to a power-law probability with an exponent $\delta$. Correlations in the time series are brought about by applying a finite rise time $\tau_r$ and a longer decay time $\tau_d$ to the flare. Thus, for example, we apply an added intensity boost of $I_0 \exp(-k/\tau_d)$ if $k < \tau_d$ to the data points lying within *k* steps after the flare point. We apply an analogous process to leading times and $\tau_r$. Thus we investigate the properties of a model for simulating intensity light curves with a sequence of random "bursts." Since the results for the 171 Å waveband in an AR core are the least affected by the noise contribution we have used them as a guide to select the following parameter values for the model. The probability of some kind of "burst" at any given step is $P_f$= 0.1, The amplitude is in the range 0.01-0.1 arbitrary units sampled from a power-law distribution with exponent $\delta = 2.1$. The impulses have an exponential rising time $\tau_r = 1.5 \min$, and an exponential decay time $\tau_d = 6 \min$. The values for the parameters have been chosen with the aim of obtaining a degree of multifractality and correlation comparable to that of the observational data.

Figure 9 presents the averaged fluctuation function (left), the corresponding Hurst exponents (center), and the averaged distribution of increments (right) for 2000 realizations of modeled light curves. The original simulations had a length of 1500 pixels. To perform the analysis, the first 150 pixels were discarded to eliminate the initial transients and match the observational data with 1350 pixels (270 minutes). For large scales, in a scaling range of 24.8 - 64.6 minutes, we find $h(2) = 1.20 \pm 0.02$ which is compatible with the numbers in Table 1 for the 171 Å signals with either positive or zero lag in the observed core region. The degree of multifractality $\Delta h = 0.74 \pm 0.04$ (or, corrected for FSE, $0.58 \pm 0.04$) is in general larger than the values obtained for the observations. The PDF of increments displayed in Figure 9 (right), which correspond to the same scales as in Figure 3, present quasi-Gaussian properties as is found for the observed data. The statistically significant excess kurtosis values (2.8, 1.9, 0.8, 0.5) are lower than those of the observed data . This implies that the data have larger contributions from outliers. Larger excess kurtosis in the model PDFs could be obtained by increasing the upper limit of the impulse amplitudes but the other parameters would have to be accordingly adjusted to be able to reproduce the data scaling properties. The skewness values (1.16, 0.91, 0.38, 0.02) are also statistically significant. The larger numbers for the smaller scales reflect the fast rise and slow decay of the impulses and differ from the data in which the skewness for the PDFs of increments is non-zero but very small.

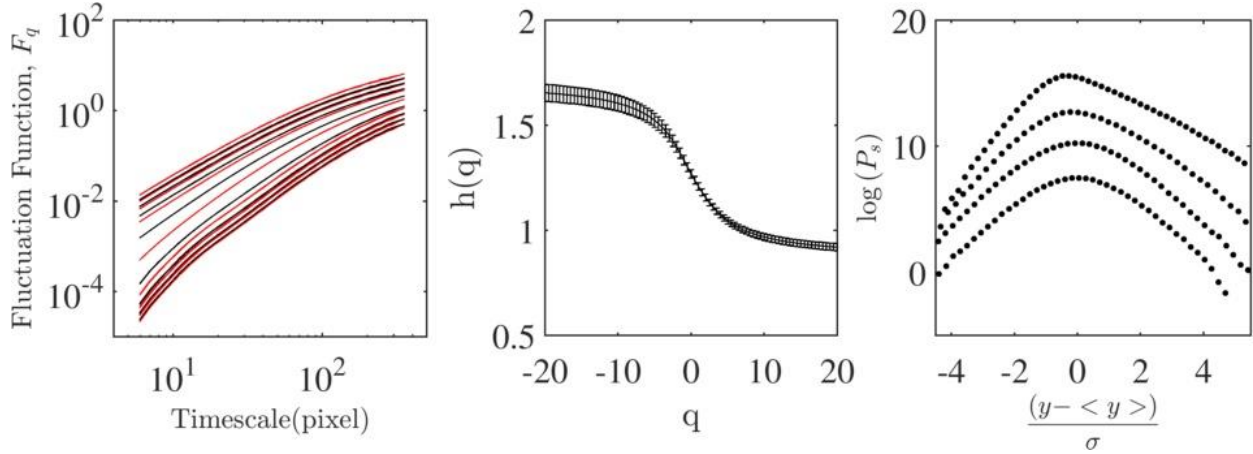

Figure 9. Left: log-log plot of the average fluctuation functions vs timescale for the modelled impulsive heating. For clearer visualization not all curves are presented. From top to bottom $q$ varies from 20 to -20 in intervals of 2.5. Center: the corresponding generalized Hurst exponents. Error bars describe the uncertainty in the linear fit to the scaling region (fitting range) in $F_q(s)$. Right: scaled distribution of increments for time scales 0.4, 1.6, 6.4 and 25.6 min in increasing order from top to bottom. To ease visualization the curves have been shifted downward in units of $\log(P_S)$ relative to the smallest scale curve. The shifts are: (1, 3, 5).

To gain more insight into how some of the model parameters can be optimized to match the observations we have investigated in detail the average fluctuation function and generalized Hurst exponent for the second moment ($F_2$). Since in the case of the model there is no excess acoustic emission which could distort the fluctuation function we can investigate the scaling for both small and large temporal scales and the crossover time where the slopes change. We have fixed $\tau_r = \tau_d/4$ and varied the decay time. We found that for large temporal scales the exponent $h(2)$ increases with increasing $\tau_d$. In the absence of noise (Figure 10 - left) the small time scales show a steep slope of $2.18 \pm 0.02$ for a scaling range 3.2 – 5.2 minutes. Although there is no direct connection to the standard Hurst exponent this value of the scaling exponent indicates the presence of strong correlations. We have calculated the autocorrelation of the time series and find for small scales an exponential decay with a time constant roughly equal to the extent $\tau = \tau_r + \tau_d$ of the duration of the impulsive burst. As noted earlier for large time scales we find an exponent of $1.20 \pm 0.02$ in a scaling range 24.8-64.6 minutes. For simulations without noise the crossover time grows or shrinks roughly with the value of $\tau$. For the choice of parameters being considered, the crossover time is ~14 minutes. We note that as shown in Figure 5 for the coronal emission in the 171 Å waveband the fluctuation function presents a steeper slope at small temporal scales as compared to large temporal scales. Since the presence of the enhanced emission from layers lower in the atmosphere distorts the scaling properties and the fitting range is very short, the precise numerical value is not informative but it suffices to keep in mind that $h(2)$ varies between 1.5 and 2. As we discuss below in order to find a closer correspondence between the phenomenological model results and the observations, the effects of the noise component in the signal must be taken into account.

To explore the effect of noise we have superposed uncorrelated Gaussian white noise with varying amplitude on the original modeled signal. By itself this noise gives a scaling index $h(2) = 0.5$ at all scales. Hu et al. (2001) have shown that for two uncorrelated signals $x(t)$ and $z(t)$ with corresponding root mean square fluctuation functions $F_2^x(s)$ and $F_2^z(s)$, the fluctuation function of the sum of the signals $x(t) + z(t)$ is given by the superposition relation

$$F_2^{x+z}(s) = \sqrt{(F_2^x(s))^2 + (F_2^z(s))^2} \tag{9}$$

The noise contribution affects primarily the small scales of the fluctuation function, and the value of the scaling exponent gradually moves closer to 0.5. When white noise with amplitude of 0.05 is added, using the same fitting ranges as in the noise free case, we find that at small temporal scales the exponent is now $0.58 \pm 0.01$ and for large time scales $1.12 \pm 0.01$ (Figure 10, right). The crossover time is 7 min and the corrected degree of multifractality at large temporal scales is $0.30 \pm 0.02$. For small scales this result is comparable to what is observed in the case of the 335 Å "hot" channel (Figure 5, right). In this case the physical signals are relatively weak, so that the effect of the noise, which, by construction, is independent of the wavelength, is more apparent than in the cooler 171 Å waveband. For large temporal scales the model with noise amplitude of 0.05 has a stronger anti-correlation and weaker degree of multifractality than those of the 335 Å zero positive lag pixels ($h(2) = 1.29 \pm 0.03$, $\Delta h = 0.20 \pm 0.06$). Interestingly the modeled results better match the large-scale results for the zero lag pixels in the 171 Å waveband ($h(2) = 1.14 \pm 0.02$, $\Delta h = 0.48 \pm 0.03$). However, due to the presence of a relatively larger noise component, the PDFs of increments for the model (not shown here) are Gaussian.

Finally, since we found for the 335 Å weak emission region that the noise component may be positively correlated, we have investigated adding this type of signal to the original simulation. As with the Gaussian noise the generalized exponent decreases at small time scales tending toward the characteristic Hurst exponent of the noise signal. The values of the Hurst exponent and amplitude of the added noise signal, which are required to match the observational results, vary depending on the different $h(2)$ of the original simulation. For the example presented here we find that for correlated noise characterized by $H = 0.55$, the amplitude can be varied as to obtain a combined signal that has $h(2) = 1.14 \pm 0.02$ at large temporal scales and $0.63 \pm 0.02$ for small temporal scales. In principle one could consider from the beginning the impulsive model with added noise and together optimize all the parameters including the Hurst exponent and the noise amplitude to match the observed scaling.

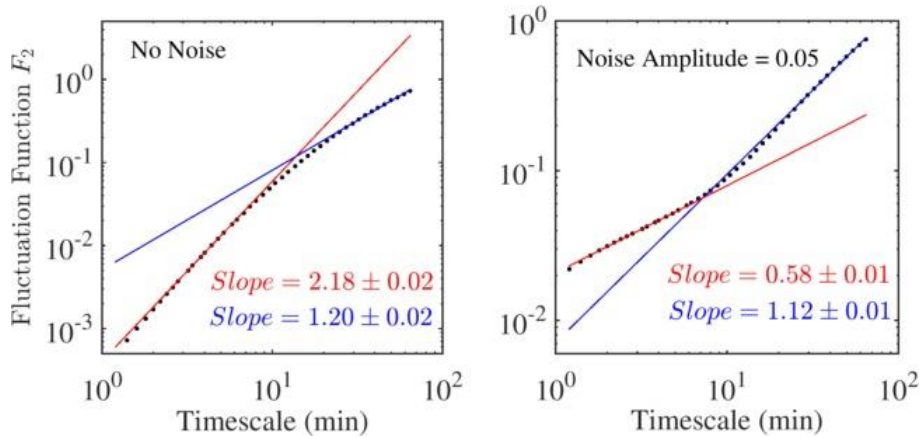

Figure 10. Left: average fluctuation function versus timescale for the modeled time series with amplitude in the range 0.01-0.1 sampled from a power law distribution with exponent $\delta = 2.1$, impulses with exponential rising time $\tau_r = 1.5$ minutes, an exponential decay time $\tau_d = 6$ min, and "flaring" probability $P_f = 0.1$. Right: average fluctuation function for the original simulation with added Gaussian white noise of amplitude 0.05.

Our goal has not been to perform a thorough evaluation of the parameter space of the model. Rather, our aim is to show that this impulsive model with added noise offers the potential to explain salient properties of the data regarding the degree of correlation in the time series, the degree of multifractality, and the statistical properties of the distribution of increments. We find that the superposition of random impulses with amplitudes satisfying a power-law distribution and with an effective degree of correlation leads to time series with multifractal properties. Addition of a signal of uncorrelated noise of varying relative strength allows us to obtain degrees of multifractality and scaling exponents $h(2)$ which approximate quantitatively those encountered in the analysis of the AIA time signals according to their relative signal-to-noise ratios. While we have not obtained meaningful quantitative values for crossover times for the AIA data to the extent that the model is applicable to explain the observations we suggest that the value of the crossover times is providing information on the average duration of the underlying heating impulses. We have found that the model also has limitations in particular regarding PDFs of the increments at varying scales. For the original run with no noise at the two smaller scales the model has much smaller kurtosis and larger skewness than those observed. The shapes of the curves also differ at these scales. As mentioned earlier one of the criteria used by Pauluhn & Solanki (2007) was to compare to the probability distribution of the values of the observed irradiances which happen to be log normal. We find that the distribution of modeled values for the choice of parameters under study best fits the distribution of intensities for the 171 Å weak emission. Even in this case the PDFs of increments at small scales do not match those of the observations. This points to the importance of combining different statistical analyses to constrain the models correctly.

## 6.2 Comparison to the Ohmic Dissipation Signal in the RMHD Model

In the previous section we showed that a phenomenological model for emitted radiative intensity can reproduce salient properties of the observations. Nevertheless these phenomenological models cannot establish a causal connection between the observed signal and coronal heating theories. Indeed the properties of the heating function are tuned ad hoc so that the emitted signal can match the observed one, but then it is very difficult to assess whether a coronal heating model would generate nanoflares with the required timescales and energy dissipation rates for the loops of interest.

Therefore in this section we report the results of applying these analysis techniques to the ohmic dissipation rate time series from a more complex three-dimensional reduced magnetohydrodynamic (RMHD) simulation of coronal loop dynamics in the attempt to use more self-consistent coronal heating models. It is expected that radiation in the hot 131 Å and 94 Å channels would be more closely fit for this analysis. As we have seen in the case of the 335 Å radiation, because the signal-to-noise ratio is small in the hot channels, the results are strongly affected by the presence of noise and the analysis is more suited to the emission in the 171 Å bandwidth. We will proceed to use the results for the 171 Å emission as reference for comparison and will expand on the validity of this choice afterward.

As suggested by Parker (1972, 1988) the shuffling of the magnetic field line footpoints by photospheric motions gives rise to a coronal magnetic field that is out of equilibrium (Rappazzo 2015). It strives to relax toward equilibrium but is continuously perturbed by the photospheric motions, therefore continuously forming current sheets thinning at the dissipative scales on ideal (Alfvénic) timescales (Rappazzo & Parker, 2013), where magnetic reconnection occurs, and leading to the impulsive release of energy identified as "nanoflares" (Parker 1988).

It was long ago shown that such a system develops nonlinear dynamics that can be seen as a magnetically dominated instance of MHD turbulence (Einaudi 1996, Dmitruk & Gómez 1997) where energy is transferred from the large scales (directly affected by convective motions) to the small scales where it is dissipated (Rappazzo & Velli 2011). As an effect of the continuous photospheric shuffling the system is characterized by a statistically steady state (Rappazzo et al. 2007), where *on average* the energy injected in the corona by the work done by photospheric motions on magnetic-field-line footpoints (the Poynting flux) is balanced by dissipation (dominated by its Ohmic component). However, although on average the energy injected in the system is fully dissipated, in time the two signals are not the same; namely there is a continuous bursty accumulation and dissipation of energy in the fashion of nanoflares (see e.g., Figure 12 in Rappazzo et al. 2008). A well-known feature of turbulence in both experiments and numerical simulations is that time series of relevant physical quantities (e.g., the total ohmic dissipation rate) become increasingly more complex, acquiring more high-frequency components, and increasingly more intermittent, for higher values of the Reynolds number (e.g., see Frisch 1995).

Clearly numerical simulations cannot implement the very high Reynolds number characteristic of coronal loop plasma, since this would imply a very high numerical resolution impossible to attain at present or in the near future. Consequently the intermittency of the simulation time series is expected to be smaller than that obtained from observations and from very high Reynolds number plasmas.

Among the different models investigated by Rappazzo et al. (2008) we consider the case (their model I) with more realistic parameters. The main characteristics of the simulation (in cartesian geometry) are: loop length 40,000 km; cross section 4,000 km (approximately four convection cells), Alfvén velocity 1000 km/s and hyperdiffusion coefficient $10^{19}$ (the use of hyperdiffusion allows one to have a higher intermittency, compatibly with the numerical resolution). Excluding the initial transient the duration of the time series is 315.4 minutes (4717 pixels), with a time step of 4.012 sec/pixel. We initially calculated the scaling properties for the time series obtained in the simulation. For a scaling range 14.0 – 41.2 minutes, close to that used in the observations, we found a generalized exponent $h(2) = 1.08 \pm 0.08$ and a corrected degree of multifractality $0.81 \pm 0.19$. In order to better compare with the observational results we have coarse-grained the time series by averaging over three temporal pixels, degrading the resolution to 12.04 s. The best fit is found for the range 13.6 – 38.5 minutes which leads to a slope of $1.15 \pm 0.04$ comparable to that encountered for the loop 2 zero-lag pixels. The corrected degree of multifractality decreases to $0.39 \pm 0.20$ which is also closer to the values obtained with the observational data. The large error bars are a consequence of the noisy fluctuation function resulting from applying the analysis to a single time series with the minimal required length. For small temporal scales the linear fit is performed over the range 2.0-4.8 minutes obtaining an exponent of $1.97 \pm 0.05$. This value indicates a positive correlation at small scales as encountered in the phenomenological model. The crossover time is 8.0 minutes. At small temporal scales the $h(q)$ curve does not show the monotonic decrease with $q$ corresponding to a multifractal. The corrected degree of multifractality is $0.21 \pm 0.32$ which together with the variations in $h(q)$ indicates that within the error bars the signal tends to be monofractal. Figure 11 (left) presents the average fluctuation function and generalized Hurst exponent of the variance ($F_2$) together with the fits at small and large scales for the coarsegrained time series. Figure 11 (center) displays the corresponding spectrum of generalized Hurst exponents for large time scales. For the shuffled time series $h(2) = H = 0.45 \pm 0.04$, which suggests that a small degree of negative correlation remains. The degree of multifractality is $\Delta h = 0.02 \pm 0.12$ which is effectively zero. For the original simulated time series the corrected degree of multifractality $\Delta h = 0.06 \pm 0.04$ is small but not zero to one sigma. Figure 11 (right) displays the scaled distribution of increments for the simulated times series at scales of (0.1, 0.5, 2.1, 8.6, 26) minutes. We present this result rather than the one for the coarse grained series since for the latter the reduced number of data points lead to very sparse histograms. The skewness values (-0.27, -0.38, -0.30, -0.16, -0.08) are all statistically significant. We note that in the case of the observations, the zero lag pixels for loop 2 presented negative skewness at all increment scales in a range of times comparable to those analyzed here. The PDF of increments is quasi-Gaussian with excess kurtosis values (1.8, 1.7, 0.4, 0.0, 0.3). The zero excess kurtosis found for the 8.6 minute increment distribution has a corresponding Z-value that indicates a normal distribution. There is no evidence of major contributions due to fat tails in the distribution of ohmic dissipation. This is compatible with the fact that in this simulation there is no accumulation of energy and therefore less opportunity for intermittent emission with large amplitudes. Finally we have investigated the effect of adding noise to the RMHD model, and find that this affects the small scale scaling as in the simple nanoflare model. For example adding Gaussian noise with an amplitude of 10% of the maximum value of the simulated Ohmic dissipation leads to a scaling exponent of $0.64\pm0.02$ in a range 0.6-2.8 minutes. The large scales, however, are minimally affected.

We have used the results from the observational data in the 171 Å bandwidth as a reference. In recent work Dahlburg et al. (2016) performed 3-dimensional fully compressible MHD simulations of a coronal loop in Cartesian geometry with setups very similar to those of Rappazzo et al. (2007, 2008). The dynamics are indeed very similar for the two cases, but the use of the energy equation (with field-aligned thermal conduction and radiative losses) in the fully compressible case allows one to compute the emitted radiation. In fact the simulated temperature and density were used to calculate emission line intensities for the EIS/Hinode instrument and in turn calculate differential emission measures (DEM). They found that the DEM from the simulations is compatible with a DEM calculated from observations characterized by a Gaussian distribution and with weighted mean temperatures corresponding to the 171 Å and 193 Å bandwidths. To the extent that the RMHD model shares similar properties to the fully compressible simulations with thermodynamics we find justification in using our data results for the 171 Å emission as a guide.

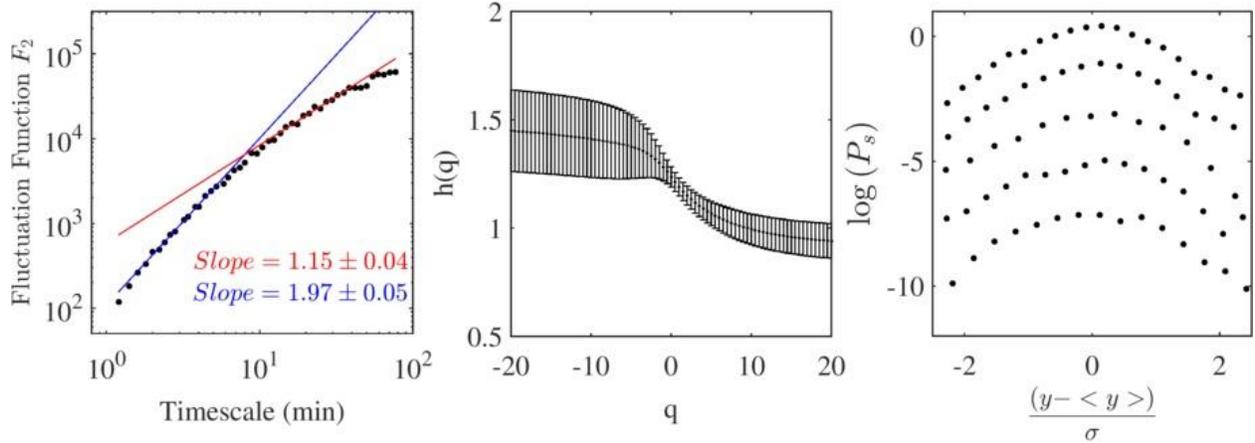

Figure 11. Left: log-log plot of the average fluctuation function $F_2$ vs timescale for the ohmic dissipation in the RHMD simulation coarse grained time series. Center: the corresponding generalized Hurst exponents. Error bars describe the uncertainty in the linear fit to the scaling region (fitting range) in the average fluctuation functions $F_q$ for general moment $q$. Right: scaled distribution of increments for the original time series. The time scales are 0.1, 1.5, 2.1, 8.6 and 26 minutes in increasing order from top to bottom. To ease visualization the curves have been shifted downward in units of $\log(P_S)$ relative to the smallest scale curve. The shifts are: (0.5, 1.7, 3, 5).

The fact that the ohmic dissipation signal exhibits multifractal properties at large temporal scales is not surprising since it results from a turbulent process. It is important to stress that even with its intrinsic approximations the analysis of the RMHD model output agrees with the results for the observational data. There is compatibility in the order of the degree of multifractality, the level of anti-persistence, and the fact that the multifractality is caused by long-term correlations. The PDFs of increments indicate that the modeled time series is less intermittent than those for the loop data, as expected due to the necessarily lower Reynolds numbers that must be implemented to properly resolve the numerical simulations.

## 7. Summary and Discussion

In this paper we have shown how, using the combined calculation of the distribution of increments and the MF-DFA of the intensity time series of EUV emissions, measured with the *SDO*/AIA, it is possible to uncover scaling properties that characterize the long-range temporal evolution of the data without making assumptions on the stationarity of the time series. By in large part eliminating macroscopic excursions in the time series we have aimed to focus the study on the background emission. The analysis was performed for three different types of regions, AR core, an adjacent area of weak emission, and two core coronal loops. A further refinement was implemented by distinguishing the pixels with positive lags from those with zero lags in both the 335-171 and 211-171 pairs. This opens the possibility to separate contributions to the observed emission from the TR from those from the corona.

The presence of noise affects the results of both types of analysis. Therefore the best information available must be obtained from the study of the 171 Å signal which has the highest signal-to-noise ratio in the collection of EUV emission wavebands. We have shown that for all regions, and at "large" temporal scales, in the approximate range of 15-45 minutes, the 171 Å emission presents a distribution of increments and variation in the generalized Hurst exponents corresponding to a multifractal. In particular the PDFs of both TR and coronal emission are "quasi-Gaussian" for large temporal scales and "leptokurtic" for small time increments, as observed in turbulent systems (eg. Tu & Marsch 1995; Bruno & Carbone 2013). In the core region the degree of multifractality (~0.3), including finite size corrections, is comparable for the TR and coronal emissions. In the case of loop 2, in which we can also compare between the types of emission, the effective degree of multifractality is larger for the TR emission (0.51) than for the corona (0.41). The degree of anti-correlation of the TR time series ($H \sim 0.15$) is stronger than for coronal emission in all physical regions ($0.26 \leq H \leq 0.41$). In all cases, at these temporal scales, the processes are anti-persistent with positive fluctuations tending to be followed by negative fluctuations and vice-versa. Except for the zero lag pixels in loop 2, the multifractality disappears when the time series elements are shuffled to a random order indicating that the multiscaling properties are a consequence of the long-term correlations in the data. For the loop 2 TR case, we find a small but non-zero degree of multifractality suggesting that in this case there is an additional contribution from values in the tail of the distribution of intensities which were not eliminated through the initial filtering. Regarding the different statistics in the TR region it is important to keep in mind that the emission is thought to be strongly influenced by the flux tube expansion (Warren et al. 2010, Tripathi et al 2010, Guarrasi et al. 2014), and in general by the topology of the magnetic field at the transition region (Kittinaradorn et al. 2009).

In the case of the 335 Å coronal emission we find that the coronal signal in the core can be described in terms of a multifractal "contaminated" by noise. The value of the Hurst exponent is compatible to that encountered in the 171 Å signal but the degree of multifractality is greatly reduced. For the weak emission region the noise component dominates and the PDFs of increments are close to Gaussians with slightly "fat" tails. The multifractality disappears and the generalized Hurst exponents appears to converge to a value of ~0.8. This would suggest that the noise signal is positively correlated differing from uncorrelated Gaussian noise.

In order to investigate whether the multifractality and correlation "signatures" found in the EUV emission can be explained in the context of the nanoflare scenario we turned to two different types of models. First we found that the simple phenomenological model of impulsive bursts, together with added noise (both uncorrelated and correlated), can be used to simulate intensity signals with long-term correlations comparable to those found in the observational data. For the choice of parameters selected, though, the degree of multifractality is larger than that encountered in the observations. Also the PDF of increments at small scales present lower excess kurtosis and very large skewness, which differ from the data. So we have demonstrated the validity of the impulsive model to explain the multiscaling properties in the data and the next step, and the subject of future work, is to thoroughly investigate the physically motivated components within it. To use the properties of the distribution of increments (shape, skewness, and kurtosis), the degree of multifractality, and the Hurst exponents to systematically analyze what are the optimal values of the nanoflare amplitude, scaling exponent, flaring frequency, duration, and shape of the heating impulses, as well as of the noise amplitude and correlation. Of interest would also be to study outputs of the hydrodynamic EBTEL model which would allow simulations which distinguish between the TR and the corona

At an entirely different level, the fact that the ohmic dissipation from the most realistic of the RMHD model simulations has the potential to explain the long-term correlation and multifractal properties of the data is highly promising. In particular the simulated time series coarse grained to the observational temporal resolution presents a correlation strength and degree of multifractality on the order of those of the observations. In contrast, for the original time series the degree of multifractality is much larger. This result illustrates how, despite the relative high temporal resolution in the *SDO* observations, we may not be measuring the full degree of intermittency in the emission time signals which would only be available at higher resolutions. The PDFs of increments differ from those of the observational data both in shape and with a much lower value of excess kurtosis. This may be a consequence of the low Reynolds number for the simulations. The analysis also shows that while the larger temporal scales present multifractality, the multiscaling is lost at the smaller temporal scales. While a proper comparison would be between the observational data and synthetic time series for 171 Å emission, this initial result is important in that in this case the detailed impulsive heating mechanism is not prescribed a priori but is determined self-consistently by the development of the turbulent nonlinear dynamics as put forward in the Parker model for coronal heating (Einaudi et al. 1996, Dmitruk & Gomez 1997, Parker & Rappazzo 2016, Velli et al. 2015).

The multifractal scaling we have found for the TR and coronal intensity fluctuations is apparent in studies of other, related, physical environments. Such scaling has recently been found in a laboratory magnetohydrodynamic (MHD) turbulence experiment (Schaffner et al. 2014, 2015).

In that work MHD turbulence in the Swarthmore Spheromak Experiment plasma was found to exhibit scale invariance and monofractal behavior at the dissipation range scales. But the fluctuations on the longer, inertial timescales were not scale invariant and instead showed multifractal behavior. The same behavior is evident in observed properties of solar wind turbulence: scale invariance and non-Gaussian monofractality at shorter dissipation timescales, and non-scale invariance and multifractality at longer inertial timescales (Kiyani, et al. 2009). It is suggested that the fast transition between the two types of scaling may be due to the generation of current sheets. The results of our present study show the multifractal part of this scaling on the longer time scales (~15 to ~45 minutes). Since we do not have access to the dissipation scales with the present data it is not yet possible to study the short timescale of the phenomenon and try to establish a universality connection between the TR/corona, on the one hand, and laboratory plasmas and the solar wind on the other.

Looking forward, these results demonstrate how the characteristic "signatures" of the observational data obtained by the combined tests of the PDF of increments and the MF-DFA provide strong constraints that can be used to systematically discriminate among models for coronal heating. This can start with the simpler phenomenological models in increasing order of complexity in the treatment of the physical mechanisms, as discussed earlier. In particular as increasingly higher resolution simulations become available thanks to advancements in computational power, it will be possible to analyze and compare with observations time series with higher intermittency resulting from a more developed MHD turbulent cascade.

**Acknowledgements**

AIA data are courtesy of NASA/*SDO* and the AIA science team. A.C. Cadavid and Y. Rivera acknowledge support from the Interdisciplinary Research Institute for the Sciences (IRIS) at California State University, Northridge. We also thank the referee, whose useful input has strengthened the paper.